\documentclass[fleqn,usenatbib]{mnras}

\usepackage{newtxtext,newtxmath}

\usepackage[T1]{fontenc}

\DeclareRobustCommand{\VAN}[3]{#2}
\let\VANthebibliography\thebibliography
\def\thebibliography{\DeclareRobustCommand{\VAN}[3]{##3}\VANthebibliography}

\usepackage{graphicx}	
\usepackage{amsmath}	
\usepackage{multirow}
\usepackage[utf8]{inputenc}
\usepackage[capitalise, nameinlink]{cleveref}
\usepackage{csquotes}

\crefname{section}{§}{§§}
\Crefname{section}{§}{§§}

\title[Lags of the kHz QPOs in XTE J1701$-$462]{Lags of the KiloHertz Quasi-Periodic Oscillations in the transient source XTE J1701$-$462}

\author[V. Peirano et al.]{
Valentina Peirano,$^{1}$\thanks{E-mail: v.peirano@astro.rug.nl}
Mariano M\'endez$^{1}$
\\
$^{1}$Kapteyn Astronomical Institute, University of Groningen, P.O. BOX 800, 9700 AV Groningen, The Netherlands\\
}

\date{Accepted XXX. Received YYY; in original form ZZZ}

\pubyear{2022}

\begin{document}
\label{firstpage}
\pagerange{\pageref{firstpage}--\pageref{lastpage}}
\maketitle

\begin{abstract}
We analysed 14 observations with kilohertz quasi-periodic oscillations (kHz QPOs) of the neutron star X-ray binary XTE J1701$-$462, the first source to show a clear transition between atoll and Z-like behaviour during a single outburst. We calculated the average cross-spectrum of both atoll and Z-phase observations of XTE J1701$-$462 between a reference/hard band (6.1 - 25.7 keV) and a subject/soft band (2.1 - 5.7 keV) to obtain, using a novel technique, the average time lags of the lower and upper kHz QPOs. During the atoll phase, we found that at the frequency of the lower kHz QPO the soft photons lag behind the hard ones by 18 $\pm$ 8 $\mu$s, whereas during the Z phase the lags are $33\pm35$ $\mu$s, consistent with zero. This difference in the lags of both phases suggests that in XTE J1701$-$462, as observed in other sources, the lags decrease with increasing luminosity. We found that for both the atoll and Z phase observations the fractional rms amplitude increases with energy up to $\sim$10 keV and remains more or less constant at higher energies. Since these changes in the variability of XTE J1701$-$462 occur within the same outburst, properties like the mass of the neutron star or the inclination of the system cannot be responsible for the differences in the timing properties of the kHz QPOs in the atoll and Z phase. Here we suggest that these differences are driven by a Comptonizing component or corona, possibly oscillating in a coupled mode with the innermost regions of the accretion disc.
\end{abstract}




\section{Introduction}
\label{sec:intro}
Kilohertz quasi-periodic oscillations (kHz QPOs), fast and highly coherent variability in the emission of a source, have been detected in neutron star low-mass X-ray binaries (LMXBs) since the first kHz QPO was observed in 1996 \citep{strohmayerMillisecondXRayVariability1996, vanderklisDiscoverySubmillisecondQuasiperiodic1996}. Among the different types of variability observed in the emission of neutron star LMXBs \citep[see][for a review]{vanderklisReviewRapidXray2004}, kHz QPOs have the highest frequency in the power density spectra (PDS), with central frequencies spanning from 250 to 1200 Hz \citep{mendezHighFrequencyVariabilityNeutronStar2021}.

Quasi-periodic oscillations in the emission of neutron star LMXBs are characterised by three basic parameters: their central frequency, $\nu_{\mathrm{central}}$, their quality factor, $Q = \nu_{\mathrm{central}}/$FWHM, where FWHM is the full width at half the maximum of the power of the QPO \citep[see][]{belloniUnifiedDescriptionTiming2002}, and their fractional rms amplitude, \textit{rms}. Usually kHz QPOs appear in pairs in the PDS and are called, respectively, upper and lower kHz QPOs, according to their relative Fourier frequency.

Studying the phenomena behind kHz QPOs has remained of great interest due to the tight relation between the short timescales of the variability and the dynamical timescales of the inner accretion flow close to neutron stars \citep{stellaCorrelationsQuasiperiodicOscillation1999, psaltisOriginQuasiPeriodicOscillations2000, psaltisModelsQuasiperiodicVariability2001}. Understanding the nature of kHz QPOs, and the mechanism that produces them, can potentially shed light onto the physics of neutron stars and the study of environments under the influence of their strong gravity \citep[see e.g.][]{millerEffectsRapidStellar1998, vanderklisQPOPhenomenon2005, psaltisProbesTestsStrongField2008}.

Multiple models have been proposed to explain the characteristics of kHz QPOs, however, none of them can represent all of the QPO properties simultaneously in a consistent way. Dynamical models that specifically try to explain the Fourier frequencies of kHz QPOs and relations among them are, for example: kHz QPOs as Keplerian oscillators under the influence of a rotating frame of reference \citep{osherovichKilohertzQuasiperiodicOscillations1999, titarchukRayleighTaylorGravityWaves2003}, the sonic-point beat-frequency model \citep{millerSonicPointModelKilohertz1998, lambSonicPointSpinResonanceModel2003}, kHz QPOs as resonances between the orbital and radial epicyclic relativistic frequencies in the accretion disc \citep{kluzniakPhysicsKHzQPOs2001} and the relativistic precession model \citep{stellaLenseThirringPrecessionQuasiperiodic1997}. Deeper studies of spectral-timing properties of kHz QPOs can provide a more complete picture of the phenomena that produce the variability and its nature. For instance, by studying the correlation between the emission in different energy ranges using higher order Fourier techniques like the cross-spectrum, it is possible to constrain the emission processes involved in the variability \citep[see][for a review of Fourier spectral timing techniques]{uttleyXRayReverberationAccreting2014}.

Energy-dependent time lags are among these higher order timing products used to study the X-ray variability in LMXBs. The time lags of kHz QPOs were first studied by \citet{vaughanDiscoveryMicrosecondTime1997} and \citet{kaaretDiscoveryMicrosecondSoft1999}. Both groups found that, at the frequency of the lower kHz QPOs in 4U 1608$-$52 and 4U 1636$-$53, respectively, the soft X-ray photons consistently lag behind the hard X-ray photons by around 25 $\mu$s\footnote{When the soft photons lag behind the hard ones, the lags are called soft. When the hard photons are the ones lagging behind the soft ones, the lags are called hard.}. Since then, spectral-timing studies have been performed on other low mass X-ray binaries to measure lags more precisely, finding soft lags at the frequency of the lower kHz QPO \citep[see e.g.][]{deavellarTimeLagsKilohertz2013, barretSOFTLAGSNEUTRON2013, troyerSpectraltimingAnalysisLower2017} and lags that are consistent with zero or slightly hard at the frequency of the upper kHz QPO \citep[see e.g.][for studies of the lags of both kHz QPOs]{deavellarTimeLagsKilohertz2013, deavellarPhaseLagsQuasiperiodic2016, peilleSpectraltimingPropertiesUpper2015, troyerSystematicSpectralTimingAnalysis2018}. The nature of the lags between the soft and hard X-ray bands at the QPO frequencies is still a subject of debate, with no clear interpretation of the phenomena causing them and their relation to the physics of the source. 

For a long time Comptonization has been considered to be the source of the hard X-ray component in the spectra of X-ray binaries \citep{thorneCygnusX1Interpretation1975, shapiroTwotemperatureAccretionDisk1976, sunyaevComptonizationXraysPlasma1980}, and the interaction of soft photons with the Comptonizing region, or corona, to be a potential mechanism responsible for the observed time lags \citep{leeComptonizationQPOOrigins1998}. In principle, inverse Compton scattering leads to hard lags, since the most energetic photons are the ones that suffered the most number of scatterings and are, therefore, the ones that emerge last from the system. Soft lags are explained in Comptonization models by considering an oscillation in the temperature of the source of the soft photons produced through feedback by an oscillation in the temperature of the corona, with a fraction of the Comptonized photons returning to the soft photon source. This effect will ultimately delay the soft photons with respect to the hard photons, producing the lag that we observe \citep{leeComptonUpscatteringModel2001}. Self-consistent models that consider the changes in temperature of the corona and the aforementioned feedback onto the soft photons source have been subsequently proposed, and are capable of explaining the energy-dependent fractional rms amplitude and time lags \citep{kumarEnergyDependentTime2014, kumarConstrainingSizeComptonizing2016, karpouzasComptonizingMediumNeutron2020}.

Neutron star LMXBs, where kHz QPOs appear, are generally divided into two classes: Atoll and Z sources \citep{hasingerTwoPatternsCorrelated1989}, depending on the path that the LMXB traces in the colour-colour diagram (CCD). The differences in the evolution of a source in the CCD are believed to be related to the mass accretion rate onto the neutron star and to the geometry of the accretion flow \citep{mendezDependenceFrequencyKilohertz1999,  doneModellingBehaviourAccretion2007}. Until 2006, with the discovery of XTE J1701$-$462 \citep{remillardNewXrayTransient2006}, it was considered that atoll and Z sources were two types of intrinsically different neutron star systems, with different spectral and timing behaviour \citep{vanstraatenAtollSourceStates2003, reigTimingPropertiesSpectral2004}. However, XTE J1701$-$462 showed both atoll-like and Z-like behaviour during one single outburst \citep{homanRossiXRayTiming2007, linSpectralStatesXTE2009, homanXTEJ17014622010} becoming the first observed system to display a clear transition between the two classes. \citet{sannaKilohertzQuasiperiodicOscillations2010} studied the properties of the kHz QPOs in both phases of XTE J1701$-$462, and found that the quality factor and the fractional rms amplitude of both the lower and the upper kHz QPOs are consistently higher, at a given QPO frequency, in the atoll than in the Z phase \citep[see also][]{barretDropCoherenceLower2011}. \citet{sannaKilohertzQuasiperiodicOscillations2010} proposed that the spectral properties of XTE J1701$-$462 suggest that many of the intrinsic differences believed to exist between atoll and Z sources (e.g. magnetic field or inclination) cannot explain the differences observed in the timing properties of kHz QPOs in both phases.

The behaviour of the lag of the lower kHz QPO in atoll sources has indeed been observed to depend on energy in LMXBs, as suggested by the Comptonization with feedback model, becoming softer with increasing energy \citep[see e.g.][]{troyerSystematicSpectralTimingAnalysis2018}. \citet{peiranoKilohertzQuasiperiodicOscillations2021} studied 8 atoll LMXBs and found that the slopes of the best-fitting linear model to the time-lag spectrum and the total rms amplitude of the lower kHz QPO exponentially decrease with increasing luminosity of the source, suggesting that the mechanism responsible for the lower kHz QPO depends on the properties of the corona. \citet{peiranoKilohertzQuasiperiodicOscillations2021} also found that for the upper kHz QPO the slope of the time-lag spectrum is consistent with zero for all sources, concluding that the upper kHz QPOs have a different origin to the lower kHz QPOs \citep[see e.g.][]{deavellarTimeLagsKilohertz2013, peilleSpectraltimingPropertiesUpper2015}. The transient source XTE J1701$-$462 offers an unprecedented perspective if studied in a similar way, considering that during the transition from Z-like to atoll-like behaviour, XTE J1701$-$462 also experimented a significant change in luminosity \citep[][]{sannaKilohertzQuasiperiodicOscillations2010}.

To date, no study of the energy-dependent lags and their dependence on luminosity has been performed for XTE J1701$-$462. In this paper we combine previous studies of this source and other atoll LMXBs, with an analysis of the spectral-timing properties of the kHz QPOs observed during the 2006$-$2007 outburst of XTE J1701$-$462. We study and compare the energy dependence of the fractional rms amplitude, intrinsic coherence and lags at the frequency of the kHz QPOs in the atoll and Z phases of the outburst of XTE J1701$-$462. We also study the dependence of the average lags on the luminosity of the source, putting this result into context with the behaviour observed in other LMXBs. In \cref{sec:obs-analysis} we describe the observations and the methods used in the analysis of the data, in \cref{sec:results} we show the results of such analysis, and in \cref{sec:discussion} we discuss the scientific implications of these results.

\section{Observations and Data Analysis}
\label{sec:obs-analysis}
\subsection{Observations}
\label{subsec:obs}
There are 866 observations in the public archive of the Rossi X-ray Timing Explorer \citep[RXTE;][]{bradtXRayTimingExplorer1993} of the source XTE J1701$-$462 collected using the Proportional Counter Array \citep[PCA;][]{jahodaCalibrationRossiXRay2006}. From these observations here we studied the 14 observations that have kHz QPOs in their power spectra. We selected these observations using the criteria described in \citet{sannaKilohertzQuasiperiodicOscillations2010}.

Following \citet{homanXTEJ17014622010}, we considered that XTE J1701$-$462 was in the Z phase of the outburst from the first time it was observed \citep{remillardNewXrayTransient2006} in January 2006, until April 2007. After this date, and until the source went into the quiescent phase, we considered XTE J1701$-$462 to be in the atoll phase of the outburst. Following this criterion there are 6 observations in the Z phase of the source (ObsIDs: 93703-01-02-04, 93703-01-02-05, 93703-01-02-08, 93703-01-02-11, 93703-01-03-00 and 93703-01-03-02) and 8 observations in the atoll phase (ObsIDs: 91442-01-07-09, 92405-01-01-02, 92405-01-01-04, 92405-01-02-03, 92405-01-02-05, 92405-01-03-05, 92405-01-40-04 and 92405-01-40-05) that show kHz QPOs. Hereafter, we will refer to the observations during the Z phase of XTE J1701$-$462 as Z observations, and to the observations during the atoll phase as atoll observations. 

\subsection{Fourier timing analysis}
\label{subsec:analysis}
To study the kHz QPOs we computed Leahy-normalised PDS of each observation, using event-mode data with at least 250 $\mu$s time resolution, covering the full energy range of the instrument (nominally $2-60$ keV). For some Z observations event-mode data were not available for the entire energy range; in these cases we used a combination of event-mode data and binned data covering the full PCA band. For each observation, we calculated PDS every 16 seconds data segments, yielding a frequency range from 0.0625 Hz to 2048 Hz. Finally, we averaged all the 16-s PDS to produce a single PDS per observation.

A multi-lorentzian model plus a constant accounting for the Poisson level has been consistently used to describe the shape of the PDS of low mass X-ray binaries \citep[e.g.][]{nowakAreThereThree2000, belloniUnifiedDescriptionTiming2002, pottschmidtLongTermVariability2003, ribeiroRelationPropertiesKilohertz2017}. This model describes well the shape of the variability present in different types of sources, but is independent of the underlying physics that cause them, making it specially fit to compare the variability behaviour without making assumptions about the nature of the mechanism that produces the oscillations. The different components of the variability in the PDS can be labelled considering the strength, width and central frequency of the Lorentzian functions used to fit them \citep[e.g.][]{vanstraatenMultiLorentzianTimingStudy2002}. We used the convention $L_\ell$ and $L_u$ to label the lower and upper kHz QPOs, respectively. For a more detailed review of different kinds of variability observed in X-ray binaries see \citet{vanderklisReviewRapidXray2004}.

We used a Lorentzian model with one or two Lorentzian functions plus a constant to fit and characterise the kHz QPOs in the PDS of observations in both the atoll and Z phase of XTE J1701$-$462. For the atoll observations we fitted the PDS between 600 and 900 Hz, rebinning the data by a factor of 64; and for the Z observations we fitted the PDS between 400 and 1200 Hz, rebinning the data by a factor of 128. Since we observed only one kHz QPO during the atoll phase, which \citet{sannaKilohertzQuasiperiodicOscillations2010} identified as the lower kHz QPO, we used only one Lorentzian function to characterise it during the fit. In the Z phase we observed two simultaneous kHz QPOs, from which we identified the one at the lowest central frequency as the lower kHz QPO and the one at the highest central frequency as the upper kHz QPO. For the Z observations we used two Lorentzian functions to describe each kHz QPO during the fit. Inspection of the PDS of observations in both the atoll and Z phases showed that the lower kHz QPOs has a higher quality factor in the atoll phase than in the Z phase, confirming the results of \citet{sannaKilohertzQuasiperiodicOscillations2010}. \textcolor{red}{}
\begin{table}
    \centering
    \begin{tabular}{lcc}
    \hline
    Band & Channel range & Energy range\\
     & & keV\\
    \hline
    Reference $H$ & 14 - 61 & 6.1 - 25.7\\
    Subject $S$ & 0 - 13 & 2.1 - 5.7\\
    \hline
    \end{tabular}
    \caption{Energy bands and channel ranges for the cross-spectra, time/phase lags and intrinsic coherence computations. The energy-channel conversion corresponds to the Epoch 5 of the instrument.}
    \label{tab:CS-bands}
\end{table}
\begin{table}
    \centering
    \begin{tabular}{cc|cc}
    \hline
    \multicolumn{2}{c}{Atoll phase} & \multicolumn{2}{c}{Z phase} \\
    \hline
    Channel range & Av. energy & Channel range & Av. energy \\
    & keV & & keV\\
    \hline
    7 - 11 & 4.2 & \multirow{2}{*}{0 - 13} & \multirow{2}{*}{3.9}\\
    12 - 15	& 6 & &\\
    16 - 21 & 8 & \multirow{3}{*}{14 - 35} & \multirow{3}{*}{10.5}\\
    22 - 25	& 10.2 & &\\
    26 - 35	& 12.7 & &\\
    36 - 41	& 16.3 & 36 - 41 & 16.3\\
    42 - 49	& 18.9 & 42 - 49	& 18.9\\
    \hline
    \end{tabular}
    \caption{Energy and channel equivalent ranges of the bands used in the calculation of the fractional rms amplitude. These ranges were selected to be as close as possible to the ones defined by \citet{ribeiroAmplitudeKilohertzQuasiperiodic2019}. The energies correspond to the Epoch 5 of the instrument.}
    \label{tab:rms-bands}
\end{table}

We found that, in the PDS of the atoll observations, the central frequency of the lower kHz QPOs varies considerably throughout the length of one entire observation. To trace these changes in central frequency, we constructed dynamical power spectra of the atoll observations using 16-s segments PDS and assigning a unique $L_\ell$ central frequency to each segment. When necessary, we combined multiple contiguous 16-s PDS until it was possible to identify a unique central frequency value for those segments combined. Using these central frequencies, we split every PDS into 8 different frequency selections, with limits in frequency listed in \cref{tab:PSfit_lag_atoll}. We shift-and-added \citep{mendezDiscoverySecondKilohertz1998} together every 16-s PDS of each frequency selection into one unique PDS, shifting the kHz QPOs to a frequency in the centre of the corresponding frequency selection.

\begin{figure}
    \centering
    \includegraphics[width=\linewidth]{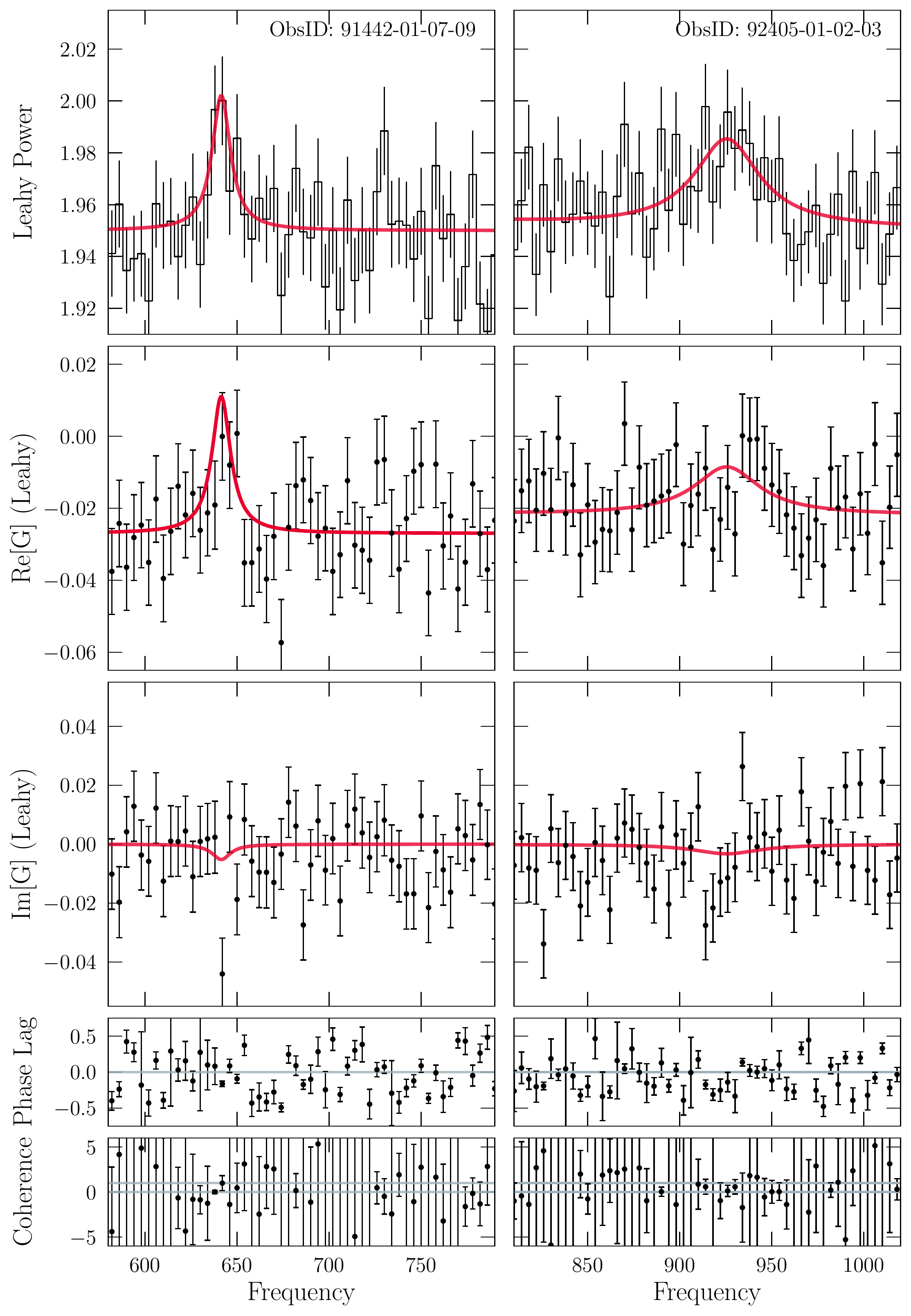}
    \caption{\textit{Top panel:} PDS of the observations 91442-01-07-09 and 92405-01-02-03 of XTE J1701$-$462 during its Z phase. The fitted Lorenztian function to the kHz QPO is shown in red on top of the histogram. \textit{Second and third panels:} Real and imaginary parts of the cross-spectrum of the two Z observations. The red solid line shows the joint best-fitting Lorentzian functions. \textit{Fourth and fifth panels:} Phase lag and intrinsic coherence as a function of Fourier frequency. In the fourth panel, the grey solid line indicates zero phase lag. In the fifth panel, the the grey solid lines indicate coherence equal to one (perfect coherence between signals) and equal to zero (completely incoherent signals).}
    \label{fig:pscross-data-Z}
\end{figure}

\begin{figure}
    \centering
    \includegraphics[width=\linewidth]{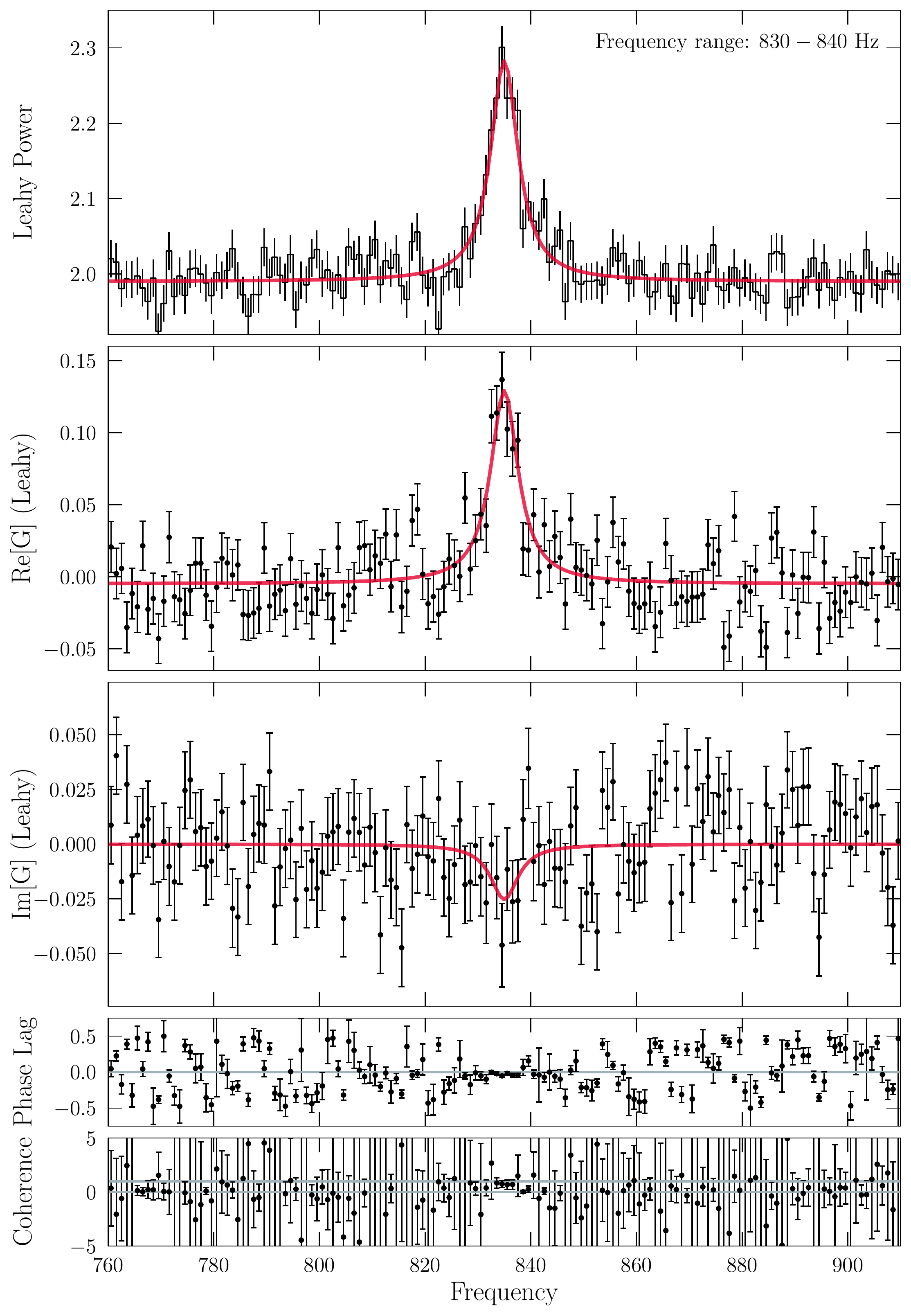}
    \caption{\textit{Top panel:} Shifted-and-added PDS of XTE J1701-462 during its atoll phase, for the observations where the frequency of the lower kHz QPO was between 830 and 840 Hz. \textit{Second and third panels:} Real and imaginary parts of the cross-spectrum of the atoll frequency selection. The red solid line shows the joint best-fitting Lorentzian function. \textit{Fourth and fifth panels:} Phase lag and intrinsic coherence as a function of Fourier frequency. In the fourth panel, the grey solid line indicates zero phase lag. In the fifth panel, the the grey solid lines indicate coherence equal to one (perfect coherence between signals) and equal to zero (completely incoherent signals).}
    \label{fig:pscross-data-atoll}
\end{figure}

In contrast, when analysing the PDS of the Z observations, because the QPOs are weaker and broader than in the atoll phase, it is not possible to significantly detect changes in the central frequency of the kHz QPOs within the length of a single observation. For this reason we analysed the PDS of every full observation during the Z phase separately (see \cref{tab:PSfit_lag_Z}). In \cref{fig:pscross-data-Z} and \cref{fig:pscross-data-atoll} we show examples of the PDS from two Z observations (91442-01-07-09, with a more significant lower kHz QPO, and 92405-01-02-03, with a more significant upper kHz QPO) and the PDS of the 830$-$840 Hz frequency selection of the atoll observations.

To calculate the average cross-spectra, $\mathrm{G}=\mathrm{Re[G]}+i\mathrm{Im[G]}$, we computed complex Fourier transforms for both the atoll frequency selections and the Z observations in two different bands: a reference (hard) band and a subject (soft) band (see \cref{tab:CS-bands} for the limits in RXTE channels and equivalent energy of the bands that correspond to the Epoch 5 of the instrument, as all our observations were performed towards the end of the RXTE mission). Equivalently to the atoll PDS procedure described in the previous paragraph, we calculated the atoll complex Fourier transforms of each frequency selection in \cref{tab:PSfit_lag_atoll} shift-and-adding (using the already defined central frequencies) 16-s segments with a time resolution of 250 $\mu$s, and a minimum and maximum Fourier frequency of 0.0625 Hz and 2048 Hz, respectively. Examples of real and imaginary parts of these averaged cross-spectra are shown in the second and third panels of \cref{fig:pscross-data-Z} and \cref{fig:pscross-data-atoll}, for the atoll and Z phase, respectively. The errors reported for both the real and imaginary parts of the cross-spectra are given by Eq. (13) in \citet{ingramErrorFormulaeEnergydependent2019}.

Using these averaged cross-spectra and following the procedure in \citet{nowakRossiXRayTiming1999}, we calculated the phase lag and intrinsic coherence as a function of the Fourier frequency for each phase (bottom panels in \cref{fig:pscross-data-Z} and \cref{fig:pscross-data-atoll}). Considering that we use the hard band as the reference band, positive values of the lags represent the soft photons lagging the hard ones, while negative values represent hard photons arriving after the soft ones.
\begin{table*}
    \centering
    \begin{tabular}{ccccccccc}
    \hline
    Z phase & \multicolumn{4}{c}{$L_\ell$} & \multicolumn{4}{c}{$L_u$} \\
    ObsID & $\nu_{\mathrm{central}}$ (Hz) & rms (\%) & FWHM (Hz) & Phase lag & $\nu_{\mathrm{central}}$ (Hz) & rms (\%) & FWHM (Hz) & Phase lag\\
    \hline
    91442-01-07-09 & $641.5 \pm 1.8$ & $1.3 \pm 0.2$ & $12.2 \pm 1.2$ & $0.39 \pm 0.34$ & $-$ & $-$ & $-$ & $-$\\
    92405-01-01-02 & $617.8 \pm 16.8$ & $2.9 \pm 0.7$ & $102.4 \pm 57.5$ & $0.11 \pm 0.54$ & $-$ & $-$ & $-$ & $-$\\
    92405-01-01-04 & $-$ & $-$ & $-$ & $-$ & $755.3 \pm 8.6$ & $3.6 \pm 0.4$ & $117.2 \pm 31.7$ & $0.37 \pm 0.20$\\
    92405-01-02-03 & $620.2 \pm 17.6$ & $3.2 \pm 0.6$ & $172.1 \pm 69.4$ & $-0.20 \pm 0.65$ & $925.5 \pm 5.0$ & $2.1 \pm 0.3$ & $43.1 \pm 16.2$ & $0.29 \pm 0.41$\\
    92405-01-02-05 & $598.6 \pm 7.2$ & $2.0 \pm 0.3$ & $67.5 \pm 23.4$ & $0.36 \pm 0.22$ & $850.2 \pm 12.6$ & $2.2 \pm 0.4$ & $110.1 \pm 45.7$ & $0.01 \pm 0.30$\\
    92405-01-03-05 & $611.2 \pm 14.9$ & $3.1 \pm 0.7$ & $116.2 \pm 54.6$ & $-0.67 \pm 0.75$ & $916.7 \pm 8.8$ & $2.9 \pm 0.5$ & $75.8 \pm 30.9$ & $-0.12 \pm 0.29$\\
    92405-01-40-04 & $651.8 \pm 9.2$ & $3.0 \pm 0.5$ & $91.5 \pm 31.8$ & $-0.44 \pm 0.29$ & $914.0 \pm 8.8$ & $3.2 \pm 0.5$ & $103.8 \pm 35.5$ & $-0.36 \pm 0.52$\\
    92405-01-40-05 & $637.6 \pm 8.1$ & $3.5 \pm 0.5$ & $94.2 \pm 28.2$ & $0.24 \pm 0.29$ & $914.0 \pm 6.6$ & $3.3 \pm 0.4$ & $78.2 \pm 22.8$ & $0.52 \pm 0.24$\\
    \hline
    Joint fit &  &  &  & $0.13 \pm 0.14$ &  &  &  & $0.24 \pm 0.14$\\
    \hline
    \end{tabular}
    \caption{Properties of the kHz QPOs detected in the Z phase of XTE J1701$-$462. The fractional rms amplitude was calculated over the full energy range of the instrument, between 2 and 60 keV. The phase lag, between the reference-hard band and the subject-soft band in \cref{tab:CS-bands}, was calculated as described in \cref{subsec:lag-technique}, first for each Z observation individually and then in a joint fit of all Z observations to obtain a unique value for the Z phase of the source. Subscript letters $\ell$ and $u$ denote lower and upper kHz QPOs, respectively.}
    \label{tab:PSfit_lag_Z}
\end{table*}
\begin{table*}
    \centering
    \begin{tabular}{ccccc}
    \hline
    Atoll phase & \multicolumn{4}{c}{$L_\ell$}\\
    Frequency selection range (Hz) & $\nu_{\mathrm{central}}$ (Hz) & rms (\%) & FWHM (Hz) & Phase lag\\
    \hline
    $600-660$ & $621.5 \pm 0.6$ & $8.3 \pm 0.9$ & $5.5 \pm 1.7$ & $-0.06 \pm 0.25$\\
    $660-700$ & $671.6 \pm 1.3$ & $9.8 \pm 1.1$ & $13.3 \pm 3.9$ & $0.02 \pm 0.15$\\
    $700-750$ & $725.3 \pm 0.4$ & $10.2 \pm 0.5$ & $9.2 \pm 1.2$ & $0.09 \pm 0.10$\\
    $750-800$ & $775.6 \pm 0.4$ & $9.8 \pm 0.8$ & $5.5 \pm 1.2$ & $0.07 \pm 0.20$\\
    $800-830$ & $815.1 \pm 0.3$ & $9.0 \pm 0.5$ & $5.5 \pm 0.9$ & $0.12 \pm 0.12$\\
    $830-840$ & $835.0 \pm 0.2$ & $8.7 \pm 0.2$ & $6.3 \pm 0.5$ & $0.18 \pm 0.06$\\
    $840-850$ & $845.1 \pm 0.3$ & $7.6 \pm 0.5$ & $5.7 \pm 1.0$ & $-0.06 \pm 0.12$\\
    $850-950$ & $900.2 \pm 0.5$ & $6.7 \pm 0.5$ & $6.8 \pm 1.4$ & $-0.19 \pm 0.14$\\
    \hline
    Joint fit &  &  &  & $0.10 \pm 0.04$\\
    \hline
    \end{tabular}
    \caption{Properties of the kHz QPOs detected in the atoll phase of XTE J1701$-$462. The fractional rms amplitude was calculated over the full energy range of the instrument, between 2 and 60 keV. The phase lag, between the reference-hard band and the subject-soft band in \cref{tab:CS-bands}, was calculated as described in \cref{subsec:lag-technique}, first for each frequency selection individually and then in a joint fit of all frequency selections to obtain a unique value for the atoll phase of the source. Subscript letter $\ell$ denotes lower kHz QPOs.}
    \label{tab:PSfit_lag_atoll}
\end{table*}
\subsection{Average phase lag of the kHz QPOs}
\label{subsec:lag-technique}
To calculate the time and phase lags of kHz QPOs it is common to select a frequency range (using, for example, the FWHM as criterion) and average the frequency-dependent real and imaginary parts of the cross-spectrum of an observation over this frequency range \citep[see e.g.][]{deavellarTimeLagsKilohertz2013,troyerSystematicSpectralTimingAnalysis2018}. This approach gives an accurate enough value when the lag is significant enough in the cross-spectrum, which is the case for the lower kHz QPO in the atoll observations of XTE J1701$-$462, but it can be a limitation when studying the Z observations, where the kHz QPOs are less significant and frequency-dependent lags are consistent with zero.

In this paper we applied a novel technique to calculate the average lags of the the lower and upper kHz QPOs of XTE J1701$-$462, considering the entire region in the PDS where the QPO is present. We computed the average lag of the kHz QPOs directly from a joint multi-Lorentzian fit to the real and imaginary parts of the cross-spectra of each phase, calculated as described in \cref{subsec:analysis}. For the Z phase, we jointly fitted the cross-spectra from all Z observations (both linked and separately) fixing the values of the central frequency and the FWHM to the corresponding best-fitting parameters that describe the kHz QPOs in the Z-phase PDS (see \cref{tab:PSfit_lag_Z}). To directly obtain the average lags from the fit, we allowed the normalisation of the real part and the phase lag to vary and computing the normalisation of the imaginary part as $\mathrm{Im[G]} = \mathrm{Re[G]} \tan(\Delta \phi)$. This procedure is justified because the power in the full band is equal to the sum of the square of the real and imaginary parts of the Fourier transform of the full band light curve, and the signals in the two bands used to calculate the cross-spectrum are highly correlated in the frequency range of the QPOs. Because this method assumes that the lags in the frequency range over which the QPO is significant are constant, our procedure takes the full QPO profile to measure the lags.

For the Z observations 91442-01-07-09 and 92405-01-01-02 we only included the lower kHz QPO in the fit, as the upper kHz QPO is not significant in the PDS. Similarly, for the observation 92405-01-01-04 we considered only the upper kHz QPO in the fit, as the lower kHz QPO is not significant in the PDS \citep[see][]{sannaKilohertzQuasiperiodicOscillations2010}. For the atoll phase we followed a similar procedure to the one of the Z phase, but we used the cross-spectra of each frequency selection, instead of the individual atoll observations, and the corresponding best-fitting parameters to the PDS (see \cref{tab:PSfit_lag_atoll}) to perform the joint fit of the real and imaginary parts of the cross-spectra.

In all the fits we considered the so-called channel cross-talk \citep[see section 2.4.2 in][]{lewinReviewQuasiperiodicOscillations1988}, a consequence of deadtime producing correlation between energy channels, by adding a constant function to the multi-Lorentzian model, that varies independently while fitting both the real and imaginary parts of the cross-spectrum. We find that the constant component of the imaginary part is consistent with zero for all fits, which is as expected since the channel cross-talk only contributes to the real part of the cross-spectrum.

The technique we used in this paper makes the lag comparison between the atoll and Z phases of XTE J1701$-$462 consistent, avoiding bias by selecting an arbitrary frequency range in the cross-spectrum related to the FWHM of the QPO, given that the FWHM of the QPO in Z and atoll phases is significantly different \citep[][]{sannaKilohertzQuasiperiodicOscillations2010}.
\subsection{Energy-dependent rms}
\label{subsec:rms}
Additionally, we calculated the PDS of both atoll and Z observations for a set of energy ranges in order to study the fractional rms spectrum of the QPO. We selected the energy ranges to be as close as possible to the ones used in \citet{ribeiroAmplitudeKilohertzQuasiperiodic2019}, where they defined the channel limits considering the drift in energy-to-channel relation of the ranges used in \citet{deavellarPhaseLagsQuasiperiodic2016}. While for the atoll observations we were able to use exactly the same energy ranges as in \citet{ribeiroAmplitudeKilohertzQuasiperiodic2019}, in the Z observations the ranges had to be adapted due to the data structure of the observations. The limits of the energy ranges used in this paper are given in \cref{tab:rms-bands}, where the equivalent channel ranges correspond to the Epoch 5 of the instrument like in \cref{tab:CS-bands}. Again these PDS were calculated using the same shift-and-add procedure described previously to obtain one single PDS per phase and energy range, with a Fourier frequency range from 0.0625 Hz to 2048 Hz. 

To calculate the fractional rms amplitude, we fitted the high-frequency region (frequencies higher than 400 Hz) of the PDS calculated in each energy band in \cref{tab:rms-bands} with either one - for the atoll phase - or two - for the Z phase - Lorentzian functions plus a constant to account for the Poisson noise. We considered the integral power obtained from the fit of the Lorentzian function as equivalent to the total power at the frequency of the kHz QPO, $P_{\mathrm{QPO}}$. The fractional rms amplitude in percent units is then given by, 
\begin{equation}
    \mathrm{rms} = 100\sqrt{\frac{P_{\mathrm{QPO}}}{C_S + C_B}}\left(\frac{C_S + C_B}{C_S}\right)\,\mathrm{\%,}
\end{equation}
where $C_S$ is the source count rate - equivalent to the total count rate minus the background count rate, $C_B$.

\begin{figure*}
    \centering
    \includegraphics[width=\textwidth]{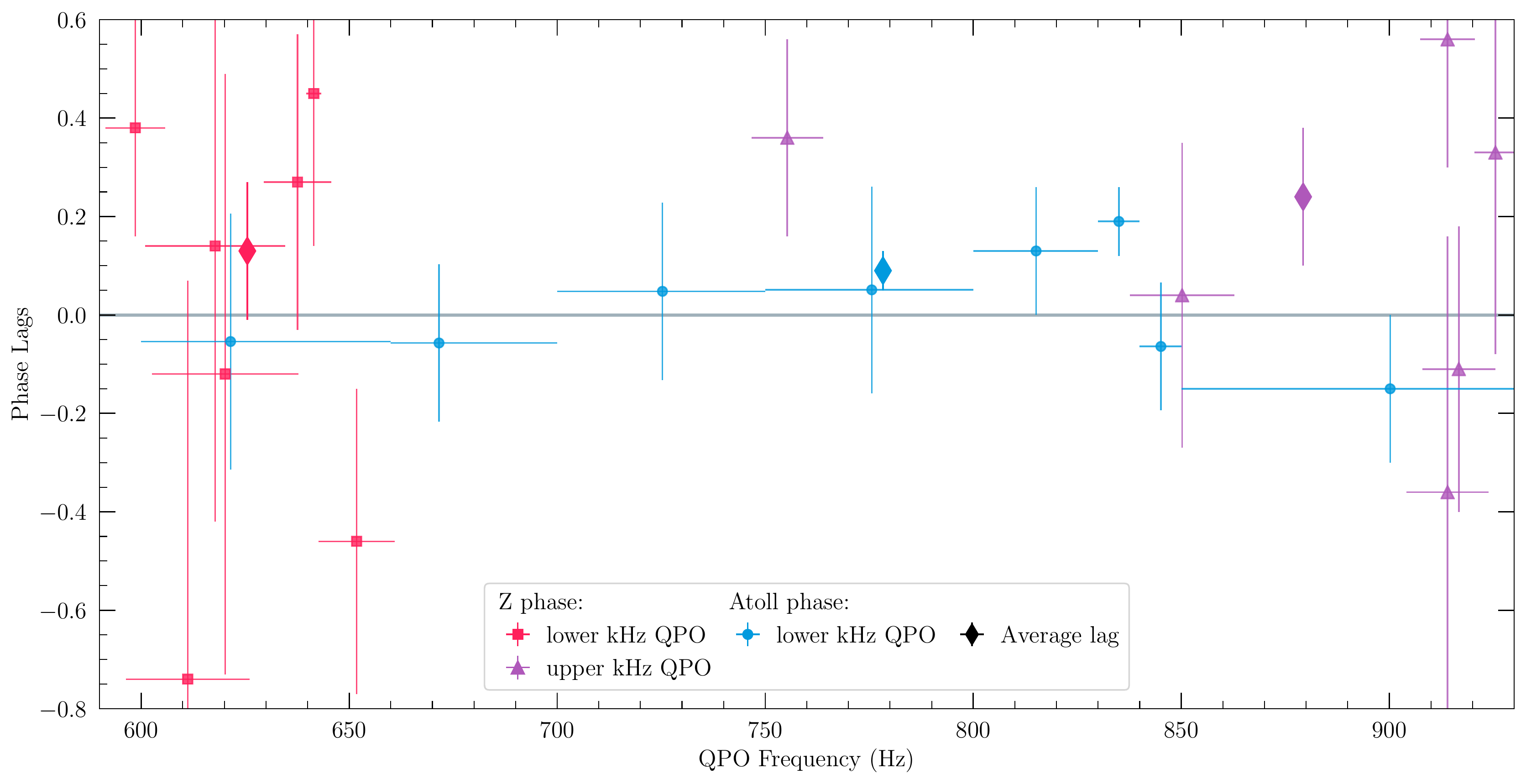}
    \caption{Phase-lag of the lower (red squares) and upper (purple triangles) kHz QPO of XTE J1701$-$462 Z observations and the lower (blue circles) kHz QPOs of XTE J1701$-$462 atoll observations as a function of QPO frequency. The error-bars of the atoll-phase lower kHz QPO lag represent the frequency bins ranges from \cref{tab:PSfit_lag_atoll}. The diamonds show the average lag value of the lower and upper kHz QPOs in the Z phase anf the lower kHz QPO in the atoll phase, with frequency equal to the mean frequency of each data group. The solid grey line indicates the zero lag.}
    \label{fig:freq_lag}
\end{figure*}
The errors reported for the rms calculations correspond to $1\sigma$, every time the kHz QPOs were detected with at least $3\sigma$ significance. In those cases in which the power of the QPO was consistent with zero or not significant enough when calculating the rms, we report the 95\% confidence upper limit, which we calculated by fixing both the central frequency and the width of the Lorentzian function to the values we obtained from the fit on the entire energy range (in \cref{tab:PSfit_lag_Z} for the Z phase and \cref{tab:PSfit_lag_atoll} for the atoll phase).
\section{Results}
\label{sec:results}
In this section we show the results of the Fourier analysis of the XTE J1701$-$462 observations during the atoll and Z phases of the outburst. In \cref{subsec:QPO-id} we show the PDS of the source during both phases and the results of the fitting procedure to identify the kHz QPOs. In \cref{subsec:phase-lags-coh} we show the analysis of the phase lags and intrinsic coherence of the signal as a function of Fourier frequency in the atoll and Z phases. In \cref{subsec:lag-QPOfreq} we examine the behaviour of the phase lags vs QPO frequency for each observation during the Z phase and for each frequency selection of the atoll phase. In \cref{subsec:avlag-lum} we study the relation between the average time lags of the lower kHz QPOs during the atoll and Z phase of XTE J1701$-$462 and the luminosity of the source, comparing our results with other atoll sources. Finally, in \cref{subsec:rms-energy} we explore the fractional rms amplitude dependence upon energy in each phase of the outburst.
\subsection{QPO identification and properties}
\label{subsec:QPO-id}
To distinguish between the lower and the upper kHz QPO, we adopted the kHz QPO identification used in \citet{sannaKilohertzQuasiperiodicOscillations2010}. In the Z phase of XTE J1701$-$462 the identification is straightforward: two peaks appear in the PDS at high frequencies, one corresponding to the lower and one to the upper kHz QPO. In contrast, during the atoll phase only one peak is significant enough in each observation, corresponding to the lower kHz QPO \citep[see figure 2 in][where a second higher frequency peak appears after applying the shift-and-add method to all atoll phase observations combined]{sannaKilohertzQuasiperiodicOscillations2010}.

In the top panels of \cref{fig:pscross-data-Z} and \cref{fig:pscross-data-atoll} we show the resulting best-fitting Lorentzian model to the PDS of two Z observations (91442-01-07-09 and 92405-01-02-03) and one frequency selection (between 830 and 840 Hz) of the atoll observations, respectively, as examples of the analysis performed over all the observations of XTE J1701$-$462. The resulting best-fitting parameters to each Z observation PDS are listed in \cref{tab:PSfit_lag_Z}, while the best-fitting parameters to each frequency selection PDS of the atoll observations are listed in \cref{tab:PSfit_lag_atoll}.
\subsection{Phase lag and coherence}
\label{subsec:phase-lags-coh}
 In \cref{tab:PSfit_lag_Z} and \cref{tab:PSfit_lag_atoll} the best-fitting values of the phase lag, calculated following the procedure described in \cref{subsec:lag-technique}, of both lower and upper kHz QPOs in the Z and atoll phases, are shown. In the atoll phase the average phase lag over all the selections of the lower kHz QPO frequency is $0.08 \pm 0.04$ radians, which means that the soft photons lag the hard ones by approximately 16 $\mu$s at this frequency. In the Z phase the average phase lag at the frequency of the lower kHz QPO is $0.05 \pm 0.15$ radians and at the frequency of the upper kHz QPO is $0.25 \pm 0.16$ radians, both consistent with zero. The best-fitting Lorentzian functions to the real and imaginary parts of the cross-spectra of two Z observations and one frequency selection of the atoll observations are shown in the middle panels of \cref{fig:pscross-data-Z} and \cref{fig:pscross-data-atoll}. Since the data from each observation in the Z phase and frequency selection in the atoll phase look very similar, we show these three examples to illustrate the analysis process.
 
In the bottom panel of \cref{fig:pscross-data-atoll} is apparent that the coherence is well constrained (small error bars) around the central frequency of the QPO in the atoll phase, within a close to symmetric frequency range around $\sim$835 Hz. Immediately outside of this frequency range the errors of the coherence increase noticeably in the plot. In the bottom panels in \cref{fig:pscross-data-Z}, while we also observe smaller errors in coherence at the central frequency of the kHz QPOs, the frequency range at which the coherence appears more constrained in the Z phase is narrower and less symmetric than it is in the atoll phase. This difference in the behaviour of the coherence of the signal, and considering that during the Z phase the kHz QPOs appear weaker and broader than during the atoll phase, makes the technique we used here to calculate the lags (see \cref{subsec:lag-technique}) especially suitable to appropriately compare the two phases of XTE J1701$-$462.

\subsection{Phase lags and QPO frequency}
\label{subsec:lag-QPOfreq}
In \cref{fig:freq_lag} we show the relation between the phase-lag of the lower and upper kHz QPOs of XTE J1701$-$462 and the QPO frequency, using the data in \cref{tab:PSfit_lag_Z} and \cref{tab:PSfit_lag_atoll}. The lags of the lower kHz QPO during the atoll phase of XTE J1701$-$462 show a marginal dependence on QPO frequency (consistent with the results of \citet{barretSOFTLAGSNEUTRON2013} and \citet{deavellarTimeLagsKilohertz2013}), however, considering the errors of the measured lags, their overall trend remains constant with increasing QPO frequency. The lags of the lower and upper kHz QPOs during the Z phase observations show no clear dependence upon QPO frequency. The diamonds in the plot show the average lags we obtained from the joint fit of the cross-spectra of each phase (as described in \cref{subsec:lag-technique}). The value of the frequency for these average lags corresponds to the mean frequency of all the data points of each kHz QPO, and is meant to be used only as a reference.
\label{subsec:avlag-lum}
\begin{figure}
    \centering
    \includegraphics[width=\columnwidth]{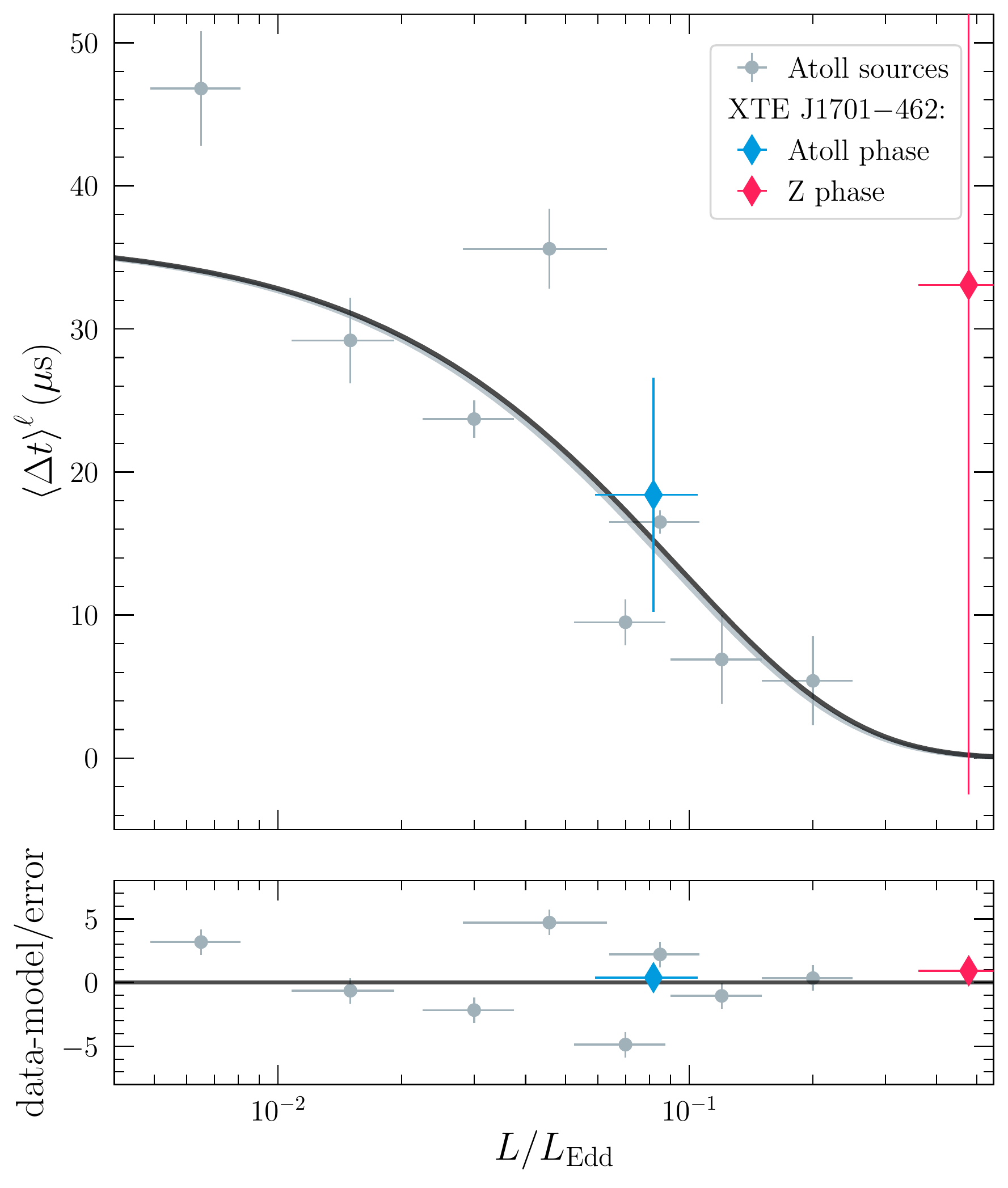}
    \caption{Average time-lag of the lower kHz QPO as a function of luminosity. The grey circles correspond the atoll LMXBs from \citet{peiranoKilohertzQuasiperiodicOscillations2021}, the blue diamond corresponds to XTE J1701$-$462 during its atoll phase and the red diamond to XTE J1701$-$462 during its Z phase. The solid black line indicates the best-fitting exponential model to the data. The solid grey line indicates the best-fitting exponential model to the atoll sample in \citet{peiranoKilohertzQuasiperiodicOscillations2021}. The value of the luminosity of XTE J1701$-$462 during its atoll and Z phases was extracted from Fig. 5 in \citet{sannaKilohertzQuasiperiodicOscillations2010}. Residuals, (data-model)/error, are also shown.}
    \label{fig:avlag-luminosity}
\end{figure}

\subsection{Average time lags and luminosity}
In \cref{fig:avlag-luminosity} we show the dependence of the average time-lag of the lower kHz QPO in the atoll and Z phases of XTE J1701$-$462 upon luminosity, and compare it to the data of the 8 atoll sources in \citet{peiranoKilohertzQuasiperiodicOscillations2021}. The solid lines in the figure represent the best-fitting exponential model, $\left\langle\Delta t\right\rangle^\ell = Ae^{-(L/L_{\mathrm{Edd}})/\alpha}$, to only the 8 atoll sources in \citet{peiranoKilohertzQuasiperiodicOscillations2021} (in \textit{grey}), with $A = 36.5 \pm 7.1$ $\mu$s and $\alpha = 0.09 \pm  0.03$; and to the same 8 atoll sources plus XTE J1701$-$462 in both phases (in \textit{black}), with $A = 36.5 \pm 6.2$ $\mu$s and $\alpha = 0.09 \pm  0.03$. From \cref{fig:avlag-luminosity} is clear that, as the luminosity increases, the average time-lags decrease exponentially for the 8 atoll sources in \citet{peiranoKilohertzQuasiperiodicOscillations2021} and XTE J1701$-$462 in the atoll phase. In the Z phase of XTE J1701$-$462 the time-lags are not as well constrained as in the atoll phase, however, the residuals show that the lags are consitent with the same trend of the atoll sources in \citet{peiranoKilohertzQuasiperiodicOscillations2021} and the atoll observations of XTE J1701$-$462.

We extracted the luminosity of XTE J1701$-$462 during its Z and atoll phases from Fig. 5 in \citet{sannaKilohertzQuasiperiodicOscillations2010}. To calculate the luminosity, \citet{sannaKilohertzQuasiperiodicOscillations2010} used a distance of 8.8 kpc, estimated by \citet{linTypeXrayBursts2009} using the Type-I X-ray bursts that occured during the 2006-2007 outburst of XTE J1701$-$462, and the $2-50$ keV flux from the source normalised by  $L_{\mathrm{Edd}}=2.5\times 10^{38}$ erg s$^{-1}$, which corresponds to the Eddington luminosity of a 1.9 M$_\odot$ neutron star accreting gas with cosmic abundance.
\begin{table}
    \centering
    \begin{tabular}{cccc}
    \hline
     & Atoll phase & \multicolumn{2}{c}{Z phase}\\
    \hline
     & $L_\ell$ & $L_\ell$ & $L_u$ \\
    \hline
    $m_1$ & 1.48 $\pm$ 0.07 & 0.33$\pm$ 0.04  & 0.38 $\pm$ 0.04\\
    $m_2$ & \multicolumn{3}{c}{fixed to 0} \\
    $E_{\mathrm{break}}$ & \multicolumn{3}{c}{9.26 $\pm$ 0.76} \\
    \hline
    \end{tabular}
    \caption{Parameters of the broken-line model fit to the fractional rms amplitude as a function of energy at the frequency of the kHz QPOs detected in both the atoll and Z phase of XTE J1701$-$462, with $E_{\mathrm{break}}$ linked and equal for all QPOs and $m_2$ fixed to zero.}
    \label{tab:linked-fixed-fit-rms}
\end{table}
\begin{figure}
    \centering
    \includegraphics[width=\columnwidth]{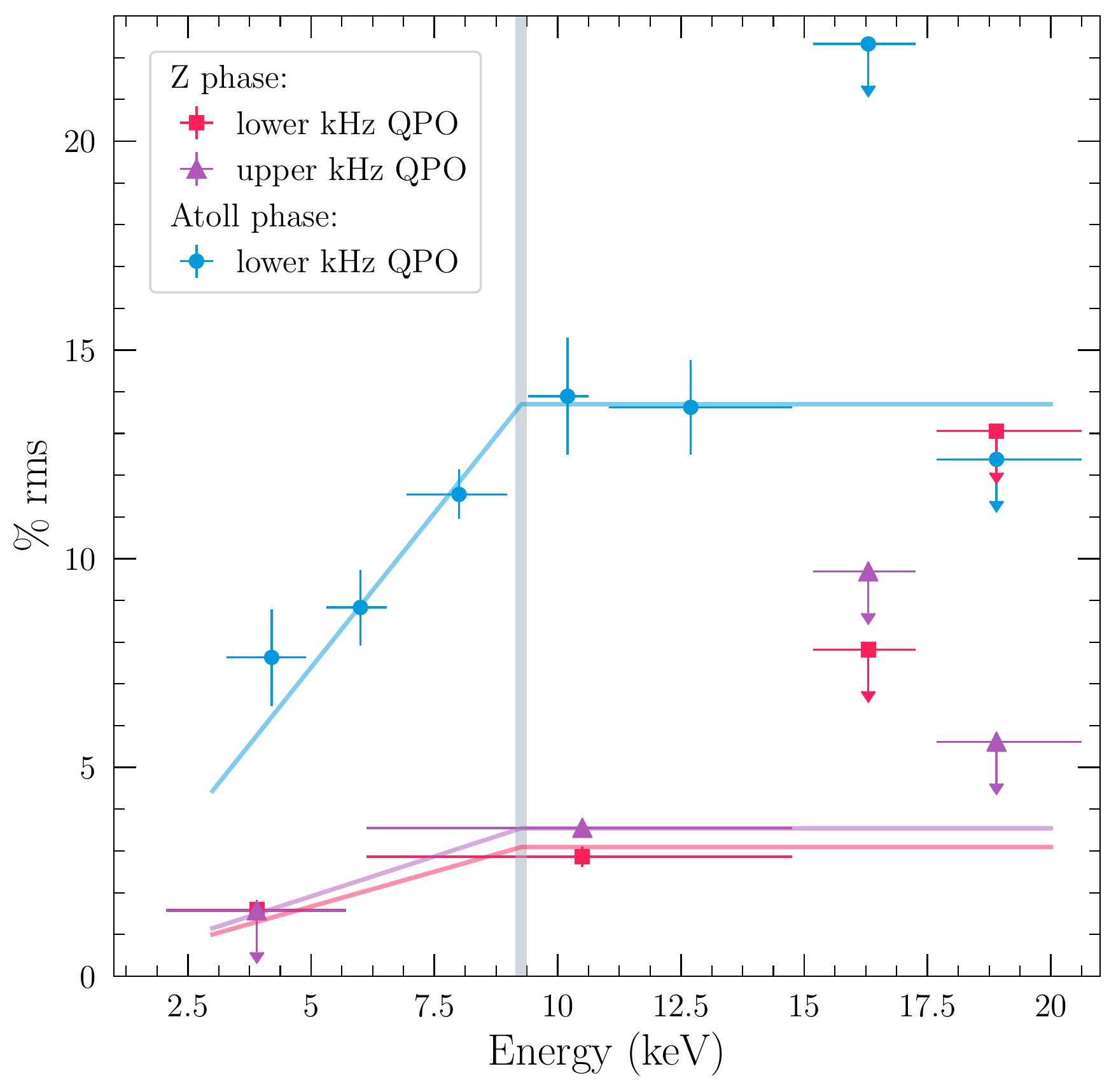}
    \caption{Fractional rms amplitude as a function of energy for the atoll phase lower kHz QPO (\textit{blue circles}) and the Z phase lower (\textit{red squares}) and upper (\textit{purple triangles}) kHz QPOs of XTE J1701$-$462. The solid lines indicate the best-fitting broken-line model, where the break energy, $E_{\mathrm{break}}$, is marked by the vertical grey line. The arrows indicate upper limits.}
    \label{fig:rmsvsenergy-allQPOs}
\end{figure}
\subsection{Fractional rms amplitude vs energy}
\label{subsec:rms-energy}
\citet{sannaKilohertzQuasiperiodicOscillations2010} studied the dependence of the fractional rms amplitude of the kHz QPOs upon the QPO frequency of XTE J1701$-$462, both in the atoll or Z phases, and showed that the fractional rms amplitude of the lower kHz QPO is consistently higher in the atoll than in the Z phase \citep[see Fig. 4][]{sannaKilohertzQuasiperiodicOscillations2010}. Similarly, \citet{sannaKilohertzQuasiperiodicOscillations2010} found that the quality factor of the lower kHz QPO is significantly higher in the atoll than in the Z phase.

In \cref{fig:rmsvsenergy-allQPOs} we show the fractional rms amplitude of both kHz QPOs as a function of energy for the atoll and Z phases, for the different energy bands defined in \cref{tab:rms-bands}. In the figure, the error-bars represent the 1$\sigma$ uncertainty and the arrows represent the 95\% confidence upper limits. In \cref{fig:rmsvsenergy-allQPOs} it is apparent that the fractional rms amplitude of the lower kHz QPO is consistently higher in the atoll than in the Z phase for all energies. In the Z phase observations, the dependence of the fractional rms amplitude upon energy for the lower and the upper kHz QPOs has no apparent significant differences. In all cases the fractional rms amplitude increases with energy up to a certain value and then remains constant for higher energies. 

To study the shape of the relation between the fractional rms amplitude and the photon energy, we fitted a broken line model to the data, given by the following equation,
\begin{equation}
\label{eq:broken-line}
   \mathrm{rms}(E)=\begin{cases}
    m_1 E & \text{, if $E<E_{\mathrm{break}}$}\\
    m_2 E + (m_1-m_2)E_{\mathrm{break}} & \text{, if $E\ge E_{\mathrm{break}}$},
  \end{cases}
\end{equation}
where $E_{\mathrm{break}}$ is the break energy (up until which the rms value increases), and $m_1$ and $m_2$ are the slopes of the lines before and after the break energy, respectively. 

We first performed a joint fit of the fractional rms amplitude data of all kHz QPOs (the lower kHz QPO in the atoll phase, and the lower and upper kHz QPOs in the Z phase) using the model given by \cref{eq:broken-line} and linking both $E_{\mathrm{break}}$ and $m_2$ for all the data-sets. This fit yielded a reduced chi-squared $\chi^2_\nu = 1.037$ for 9 degrees of freedom. Since the resulting best-fitting value of $m_2 = 0.27 \pm 0.13$ was not significantly different from zero, we performed a new fit linking only $E_{\mathrm{break}}$ and fixing $m_2$ to zero for all data-sets. This new fit yielded a reduced chi-squared $\chi^2_\nu = 1.400$ for 10 degrees of freedom. The F-test between the broken-line model with $E_{\mathrm{break}}$ and $m_2$ linked and the broken-line model with only $E_{\mathrm{break}}$ linked and $m_2 = 0$ gives a probability of 0.06. In other words, there is no statistically significant advantage in considering the slope after the break $m_2$ different from zero in the fit.

The best-fitting parameters of the fit with $E_{\mathrm{break}}$ linked and $m_2$ fixed to zero are given in \cref{tab:linked-fixed-fit-rms} and the resulting fit is shown in \cref{fig:rmsvsenergy-allQPOs} for both the Z and atoll phases. From this figure, and the values in \cref{tab:linked-fixed-fit-rms}, we found that the difference between the slope before the break $m_1$ of the lower kHz QPO in the atoll and Z phases has a significance of $\sim14\sigma$, which means that the dependence upon energy of the fractional rms amplitude of the lower kHz QPO is significantly different for both phases. Similarly, we contrasted the slope before the break for the lower and upper kHz QPOs in the Z phase only, finding that there is no significant difference between them.

In the fit of the fractional rms amplitude vs energy we used the 1$\sigma$ error as the uncertainty of the variable to fit. When the fitted Lorentzian in the corresponding energy band PDS yielded a negative integral power of the QPO in comparison with the Poisson level, we fixed the fractional rms amplitude to zero and we considered the uncertainty to be the 95\% confidence upper limit to perform the fit. In the cases where the integral power of the QPO was positive, but not significantly different from zero, we considered the original fitted value of the integral QPO power and its 1$\sigma$ error as valid during the fit.

The binning in time we performed when calculating the PDS to have a Nyquist frequency of 2048 Hz reduces the variability dominantly at high frequencies \citep[see section 4.3][]{vanderklisFourierTechniquesXray1989}, which can be relevant for the frequency ranges at which kHz QPOs are observed. Equation 4.7 in \citet{vanderklisFourierTechniquesXray1989} gives the correction factor for this effect as follows,
\begin{equation}
\label{eq:correction-factor-beta}
   \beta = \frac{\pi\nu T/N}{\sin{\pi\nu T/N}} = \frac{\pi\nu/2\nu_{\mathrm{Nyq}}}{\sin{\pi\nu/2\nu_{\mathrm{Nyq}}}},
\end{equation}
where $\nu$ is the frequency of the QPO, $T$ is the length of the observation, $N$ is the number of power spectra binned together when calculating the PDS and $\nu_{\mathrm{Nyq}} = 1/2\Delta T$ is the Nyquist frequency of the PDS, with $\Delta T = T/N$, the length of each data segment. The correction factor, for $\nu_{\mathrm{Nyq}}=2048$ Hz, of the lower and upper kHz QPO in the Z observations is $\sim$1.04 and $\sim$1.09, respectively, and of the lower kHz QPO in the atoll observations is $\sim$1.07. These values are all within the errors reported in \cref{fig:rmsvsenergy-allQPOs}, therefore not affecting the results we present here.

\section{Discussion}
\label{sec:discussion}
We studied the timing properties of the kHz QPOs of the transient neutron-star LMXB XTE J1701$-$462 and characterised, for the first time, the frequency dependent QPO lags simultaneously for the atoll and Z phases of the source. We calculated, using a novel technique, the average lags at the frequency of the QPO both in the Z and atoll phases and discovered that during the atoll phase the time lags of the lower kHz QPO are soft, with the soft photons lagging the hard ones by around 16 $\mu$s. During the Z phase, the lags of both the lower and upper kHz QPOs are consistent with zero. We also found that the intrinsic coherence of the signal is more well constrain at the frequency of the lower kHz QPO in the atoll than in the Z phase.

Additionally, we explored the behaviour of the phase lags at different QPO frequencies and observed that while the lags of the lower kHz QPO in the atoll observations of XTE J1701$-$462 have a slight dependence upon QPO frequency, the phase lags of both the lower and upper kHz QPOs in the Z observations do not. We also studied the dependence of the average time-lags of the lower kHz QPO upon luminosity during both the Z and atoll phases. We found that the average lags follow the same trend with luminosity of other atoll neutron-star systems, with the lags decreasing exponentially with increasing luminosity. Finally, we studied the fractional rms amplitude dependence upon energy of the lower and upper kHz QPOs in each phase of XTE J1701$-$462 and discovered that, as observed in other LMXBs, the fractional rms amplitude of the lower kHz QPO increases with energy up to approximately 10 keV and then remains constant at higher energies.

\subsection{Soft lags and the corona}
\label{subsec:model}
During the atoll phase of XTE J1701$-$462 we observe more clearly that, in the presence of the lower kHz QPO, the photons in the soft band, between 2.1 and 5.7 keV, lag behind the photons in the hard band, between 6.1 and 25.7 keV. Inverse Compton scattering with feedback onto the soft photon source \citep[see][]{leeComptonUpscatteringModel2001,kumarConstrainingSizeComptonizing2016,karpouzasComptonizingMediumNeutron2020} can explain these soft lags in a scenario where it is not the disc, but the corona that is responsible for the variability we observe. Indeed, the corona dominates the X-ray emission at high energies, where we observe the variability reaching its maximum amplitude (see \cref{fig:rmsvsenergy-allQPOs}, where the fractional rms amplitude for the lower kHz QPO in the atoll phase of XTE J1701$-$462 reaches its maximum at around 10 keV).

Inverse Compton scattering occurs in the corona when electrons transfer energy to photons coming from the soft energy source in an LMXB (either the disc or the surface of the neutron star), producing a delay in the emission of hard photons, with higher energies. A fraction of these Comptonized photons return to the disc and are emitted again at lower energies and later times, producing the soft lags we observe \citep{leeComptonUpscatteringModel2001,karpouzasComptonizingMediumNeutron2020}. This model not only explains the soft lags we see in the lower kHz QPOs in our data, but also can describe the difference in sign of the lags of the upper kHz QPOs in other sources \citep[see e.g.][]{deavellarTimeLagsKilohertz2013, peilleSpectraltimingPropertiesUpper2015} and the behaviour of the lags of other variability components \citep[see e.g.][]{miyamotoDelayedHardXrays1988, fordMeasurementHardLags1999}.

\subsection{Variability in the Z and atoll phases}
We observe multiple differences between the variability properties of the kHz QPOs in the Z and atoll phases of XTE J1701$-$462. More evident in \cref{fig:rmsvsenergy-allQPOs} is the consistently higher fractional rms amplitude of the lower kHz QPO in the atoll phase of the source when compared with the lower and upper kHz QPOs present in the Z phase. This discrepancy in the timing properties of the kHz QPOs in both phases of XTE J1701$-$462 are comparable with observed differences in the variability of atoll and Z LMXBs. For example, \citet{mendezMaximumAmplitudeCoherence2006} studied 12 LMXBs and found that in the Z sources of his sample the maximum rms amplitude of both kHz QPOs and the maximum quality factor of the lower kHz QPO are consistently lower than in the atoll sources.

In \cref{fig:pscross-data-Z} and \cref{fig:pscross-data-atoll}, we also observe a difference in the intrinsic coherence of the signal at the frequency of the QPOs, between both phases. In the figures it is clear that the frequency range within which the uncertainty of the intrinsic coherence is smaller, is not as well constrained in the Z than in the atoll observations. During the atoll phase of the source, the intrinsic coherence remains visibly more stable when the QPO is present in the PDS. The difference in the intrinsic coherence of the signal, although consistent with other LMXBs \citep[see e.g.][]{barretDropCoherenceLower2011}, makes using it as a criterion to measure the average lag over a frequency range, and compare both phases consistently, difficult. Here we used a novel technique to calculate the average lags at the frequency of the QPO (see \cref{subsec:lag-technique}), that allows for a more consistent comparison between the atoll and Z phases. In this method we use the Lorentzian function that describes the kHz QPOs in the PDS to find the amplitudes of the real and imaginary parts of the cross-spectra, without averaging them over a certain arbitrary frequency range defined by the coherence of the signal (which will, particularly in the case of the Z observations, lead to inconsistencies in the obtained lags).

We also studied the behaviour of the phase lags of the lower and upper kHz QPOs with respect to the QPO frequency, for both the atoll and Z phases. In \cref{fig:freq_lag} we observe a weak dependence upon QPO frequency of the lags of the lower kHz QPO in the atoll phase. Despite the fact that the trend of these lags is also consistent with a constant model, the slight increase and then decrease with frequency we see in \cref{fig:freq_lag} is consistent within errors with the relations shown by \citet{barretSOFTLAGSNEUTRON2013} and \citet{deavellarTimeLagsKilohertz2013} for the lags of the lower kHz QPOs in 4U 1608$-$522 and 4U 1636$-$53, respectively. The phase lags of the lower and upper kHz QPOs in the Z phase do not show a trend with QPO frequency in \cref{fig:freq_lag}, which is also in agreement with previous studies where it has been shown that the lags of the upper kHz QPO are constant with QPO frequency \citep[see e.g.][]{deavellarTimeLagsKilohertz2013,peilleSpectraltimingPropertiesUpper2015}.

The large error of the average lags of the Z observations that we obtained using the technique described in \cref{subsec:lag-technique}, together with the differences in the behaviour of the intrinsic coherence of the signal in the Z and atoll phases and the lower fractional rms amplitude of the variability in the Z phase (already observed by \citet{sannaKilohertzQuasiperiodicOscillations2010}), suggest that the mechanism responsible for the kHz QPOs is affected during the transition from a Z-like source to an atoll-like source of XTE J1701$-$462.

Differences in the properties of kHz QPOs of atoll and Z sources, like the ones we observe before and after the transition of XTE J1701$-$462, are believed to be related to the geometry of the accretion flow and the mass accretion rate of the source \citep[see][for a more detailed discussion]{mendezMaximumAmplitudeCoherence2006}. \citet{sannaKilohertzQuasiperiodicOscillations2010} discarded changes in the magnetic field, neutron-star mass and inclination of the system as responsible for the variations we observe in the high-frequency variability in our data, as these changes cannot occur within the time-scale of the transition of XTE J1701$-$462. Furthermore, \citet{sannaKilohertzQuasiperiodicOscillations2010} studied how the maximum quality factor and rms amplitude of the lower kHz QPOs in both the Z and atoll phases of XTE J1701$-$462 depended upon the luminosity of the source and found that this relation follows the trend found by \citet{mendezMaximumAmplitudeCoherence2006} for a sample of 12 atoll and Z LMXBs. These results characterise XTE J1701$-$462 as a unique case to study, as the behaviour of its variability follows the relations observed in other sources, while many of the fundamental properties of the source remain constant.

\subsection{Lags and the atoll- and Z-phase luminosity}
\citet{peiranoKilohertzQuasiperiodicOscillations2021} studied the relation between the slope of the time-lag spectrum, $m$, and the luminosity of the source for eight atoll LMXBs and found that $m$ decreases exponentially with increasing luminosity. These authors found a similar relation between the average time-lags and the luminosity of the source. Considering these results, it is interesting to check whether the relation of the average time-lags and luminosity holds for the Z and atoll phases of XTE J1701$-$462.

In \cref{fig:avlag-luminosity} we combine the data from \citet{peiranoKilohertzQuasiperiodicOscillations2021} with the data of XTE J1701$-$462, using the luminosities for the Z and atoll phases of XTE J1701$-$462 given by \citet{sannaKilohertzQuasiperiodicOscillations2010}. In the top panel of the figure it is apparent that XTE J1701$-$462, during its atoll phase, follows the trend of the other atoll sources very closely. Indeed, the black solid line, that represents the best-fitting exponential model to all data points, lies almost exactly over the grey solid line that represents the best-fitting exponential model to only the atoll sources in \citet{peiranoKilohertzQuasiperiodicOscillations2021}. During the Z phase the average lags have a large uncertainty, however, in the bottom panel of \cref{fig:avlag-luminosity} it is apparent that during the Z phase of XTE J1701$-$462 the relation between the average lags of the lower kHz QPO and luminosity also holds.

\citet{peiranoKilohertzQuasiperiodicOscillations2021} suggested that the similar relations between the slope of the time-lag spectrum and the total rms amplitude with the luminosity of the source (both decrease exponentially with increasing luminosity) imply that there is one single property of the system in these LMXBs that drives the behaviour of the variability. Since we observe a similar dependence upon luminosity of the average lags in both Z and atoll phases of XTE J1701$-$462, we can conclude that a similar effect is taking place in this source while it transitioned from one phase to the other. This relation suggests that it is the corona that is responsible for the changes we observe in the variability, as its contribution in the energy spectrum changes with luminosity\footnote{We use here the luminosity as a proxy for the properties of the corona as done in \citet{peiranoKilohertzQuasiperiodicOscillations2021}.}. This scenario fits with the inverse Compton scattering model with feedback onto the soft photon source \citep[see][]{leeComptonUpscatteringModel2001,kumarConstrainingSizeComptonizing2016,karpouzasComptonizingMediumNeutron2020} we described in \cref{subsec:model}, as in this model is also the corona the one that drives the properties of the variability. Our results confirm what was suggested in \citet{peiranoKilohertzQuasiperiodicOscillations2021}, where they described the mechanism that modulates the kHz QPOs as a coupled mode of oscillation between the corona and the disc. In this \enquote{coupled oscillation mode}, changes in the luminosity - if used as a proxy of the properties of the corona - when XTE J1701$-$462 transitions from the Z to the atoll-like behaviour, the fractional rms amplitude and the lags of the kHz QPOs should decrease, exactly as we observe in our data.

Studies of the behaviour of time and phase lags with luminosity for Z neutron star LMXBs, that are located in the high-luminosity end of \cref{fig:avlag-luminosity}, could help elucidate the real nature of the mechanism responsible for the high-frequency variability we observe. To date no systematic study of the timing properties of Z sources has been performed, but the results shown in the present paper suggest that the average lags of the lower kHz QPO should decrease exponentially as the luminosity increases. A more in depth analysis of the results presented in this paper, through the glass of the model described in \citet{karpouzasComptonizingMediumNeutron2020} and \citet{garciaTwocomponentComptonizationModel2021} could also help understand the changes in the geometry of the accretion flow that can be driving the changes in the properties of the variability when XTE J1701$-$462 is transitioning from the Z to the atoll phase.



\section*{Acknowledgements}

The authors wish to thank Federico García and Kevin Alabarta for useful discussions that helped develop the data analysis methods used in this paper. They also thank the referee for constructive comments that helped improve the manuscript. This research has made use of data and software provided by the High Energy Astrophysics Science Archive Research Center (HEASARC), which is a service of the Astrophysics Science Division at NASA/GSFC.

\section*{Data Availability}
The data underlying this article are publicly available at the website of the High Energy Astrophysics Science Archive Research Center (HEASARC, \url{https://heasarc.gsfc.nasa.gov/}).


\bibliographystyle{mnras}
\bibliography{references}







\bsp	
\label{lastpage}
\end{document}